\documentclass{llncs}
\usepackage{yfonts}
\usepackage{amssymb}
\usepackage{units,stmaryrd}
\usepackage{graphicx}
\usepackage[all]{xy}
\usepackage{xspace}
\usepackage{makeidx}

\begin{document}


\mainmatter              
\title{Complexity fits the fittest}
\titlerunning{Complexity fits the fittest}  
%
\author{Joost J. Joosten}
\authorrunning{J. J. Joosten} 
%
\tocauthor{J. J. Joosten}
\institute{University of Barcelona\\ 
Department of Logic, History and Philosophy of Science\\
Carrer Montalegre 6, 08001, Barcelona, Spain\\
\email{jjoosten@ub.edu}\\
\texttt{http://www.phil.uu.nl/\homedir jjoosten/}
}

\maketitle              

\begin{abstract}
In this paper we shall relate computational complexity to the principle of natural selection.
We shall do this by giving a philosophical account of complexity versus universality. 

It seems sustainable to equate universal systems to complex systems or at least to \emph{potentially complex systems}. Post's problem on the existence of (natural) intermediate degrees (between decidable and universal $\Sigma_1^0$) then finds its analog in the Principle of Computional Equivalence ($\mathbf{PCE}$). 

In this paper we address possible driving forces --if any-- behind $\mathbf{PCE}$. Both the natural aspects as well as the cognitive ones are investigated. We postulate a principle $\mathbf{GNS}$ that we call the \textit{Generalized Natural Selection} principle that together with the Church-Turing thesis is seen to be in close correspondence to a weak version of $\mathbf{PCE}$.

Next, we view our cognitive toolkit in an evolutionary light and postulate a principle in analogy with Fodor's language principle. 

In the final part of the paper we reflect on ways to provide circumstantial evidence for $\mathbf{GNS}$ by means of theorems, experiments or, simulations.
\end{abstract}


\section{Complexity and computation}\label{section:ComplexityAndComputation}

It is a standard definition in the literature to call a computational process $\Pi$ \emph{universal} if it can simulate any other computational process $\Theta$. In other words, $\Pi$ is universal if (see for example \cite{Cooper} or any other basic text book on computability theory) for any other computational process $\Theta$, we can find an easy coding protocol $\mathcal C$ and decoding protocol $\mathcal C^{-1}$ so that we can encode any input $x$ for $\Theta$ as an input $\mathcal C(x)$ for $\Pi$ so that after $\Pi$ has performed its computation we can decode the answer $\Pi ({\mathcal{C}(x)})$ to the answer that $\Theta$ would have given us. In symbols: $\mathcal{C}^{-1}(\Pi (\mathcal{C}(x)))=\Theta(x)$.

One can formalize what it means for a protocol to be easy but for the sake of this presentation that is not too relevant. Thus, if a process is universal, it can mimic all other processes if we just prepare the right input for it. It is certainly part of our intuition that complex systems can incorporate, mimic, or use, less complex systems. In this light it seems sustainable to \emph{define} complex systems as those systems that are universal. Note that under this definition a complex system need not necessarily manifest itself in a complex appearance: a universal process can mimic \emph{any} other process whence also the very easy ones. 

\section{Intermediate degrees}

In this section we study the complexity that falls in between decidable and universal in a sense to be specified below.

\subsection{Turing degrees}

For sets of natural numbers, the notion of universality can also be defined. Contrary to real-world computations, for sets of natural numbers there are infinitely many ever-increasing notions of universality. The one that corresponds to the computational notion is that of $\Sigma^0_1$ universality. A set $K$ is called $\Sigma_1^0$-universal if for any computably enumerable set $X$ (that is a set whose values we can computably enumerate but not necessarily decide for each number if it is in the set or not) there is a computable function $f_X: \mathbb{N} \to \mathbb{N}$ so that
\[
x\in X \ \ \Longleftrightarrow \ \ f_X(x) \in K .
\]
We call such a function $f_X$ also a reduction. Post \cite{Post1944} raised the famous question of whether there is some natural computably enumerable set of natural numbers that is computationally more informative than a decidable set, but less informative than the universal set $K$. 

Often, instead of speaking of sets directly one considers \emph{degrees} also called \emph{Turing degrees}. A Turing degree can be considered as the entity of all the sets that contain the same amount of information in the sense of the above considered reduction. Thus, two sets $X$ and $Y$ fall in the same degree --we write $X\sim Y$-- whenever there is some computable $f: X\to Y$ such that $x\in X \ \Leftrightarrow \ f(x) \in Y$ and some computable $g: Y \to X$ such that $y\in Y \ \Leftrightarrow \ g(y)\in X$. For two Turing degrees $X$ and $Y$ we write $X<Y$ to indicate that there is some computable $f: X\to Y$ such that $x\in X \ \Leftrightarrow \ f(x) \in Y$ but no computable $g: Y \to X$ such that $y\in Y \ \Leftrightarrow \ g(y)\in X$.

It is common practice to denote the degree of decidable sets by $\emptyset$ and the degree of $\Sigma_1^0$-universal sets by $\emptyset'$. 
Post's question stated in terms of degrees now translates to whether there exists some degree $X$ which falls strictly in between $\emptyset$ and $\emptyset'$ in terms of the above defined reduction, that is $\emptyset < X < \emptyset'$.
It took the scientific community twelve years to find such an intermediate degree. However, it is generally held that this solution does not provide a \emph{natural} intermediate degree. 

Clearly the notion of being natural is rather vague and auto-determined by the scientific community itself. A clear indication for a mathematical notion to be natural is that it occurs in various other fields as well. Likewise, applicability to other kind of problems or admitting different proof methods are typically also considered an indication of naturalness. The canonical way of finding intermediate degrees is by what are called priority arguments with finite injury and it is generally held that they do not meet the above mentioned indications for being natural. We refer the reader to \cite{Sutner2011} for a more detailed account of priority arguments in the context of this paper.

\subsection{Church-Turing thesis and $\mathbf{PCE}$}

Post's question on intermediate degrees finds it real-world analog in the Principle of Computational Equivalence ($\mathbf{PCE}$) which was postulated by Wolfram in his NKS book \cite{NKS}:

\begin{quote}
{\bf $\mathbf{PCE}$}: Almost all processes that are not obviously simple can be viewed as computations of equivalent and maximal sophistication. \index{Principle of Computational Equivalence}
\end{quote}

The processes here referred to are processes that occur in nature, or at least, processes that could in principle be implemented in nature. 
Thus, processes that require some oracle or black box that give the correct answer to some hard questions are of course not allowed here.

As noted in the book, $\mathbf{PCE}$ implies the famous Church-Turing Thesis (again, see \cite{Cooper} for more details)\index{Church-Turing Thesis} ($\mathbf{CT}$):

\begin{quote}
{\bf $\mathbf{CT}$}: Everything that is algorithmically computable is computable by a Turing Machine.
\end{quote}

Both theses --$\mathbf{PCE}$ and $\mathbf{CT}$-- have some inherent vagueness in that they try to capture/define an intuitive notion. While the $\mathbf{CT}$ thesis aims at defining the intuitive notion of algorithmic computability, $\mathbf{PCE}$ aims at defining what degrees of complexity occur in natural processes. But note, this is not a mere definition as, for example, the notion of what is algorithmically computable comes with a clear intuitive meaning. And thus, the thesis applies to all such systems that fall under our intuitive meaning. 

As a consequence, the $\mathbf{CT}$ thesis would become false if some scientists were to point out an algorithmic computation that cannot be performed on a Turing Machine with unlimited time and space resources. With the development and progress of scientific discovery the thesis has to be questioned and tested time and again. And this is actually what we have seen over the past decades with the invention and systematic study of new computational paradigms like DNA computing \cite{DNAcomputing}, quantum computing \cite{QuantumComputing}, membrane computing \cite{MembraneComputing},  etc. Most scientists still adhere to the $\mathbf{CT}$ thesis. 

But the $\mathbf{PCE}$ says more. It says that the space of possible degrees of computational sophistication between obviously simple and universal is practically 
void.
In what follows we shall address the question what might cause this. We put forward two observations. First we formulate a natural candidate principle that can account for $\mathbf{PCE}$ and argue for its plausibility. Second, we shall briefly address how cognition can be important. In particular, the way we perceive, interpret and analyze our environment could be such that in a natural way it will not focus on intermediate degrees even if they were there.

\section{Complexity and Evolution}\label{section:ComplexityAndEvolution}
In this section we shall dwell on the intimate relation between evolution and the emergence of complexity. We shall follow \cite{JoostenGNS} in great lines citing certain passages but also adding new insights.

In various contexts but in particular in evolutionary processes one employs the principle of Natural Selection,  often also referred to as Survival of the Fittest. These days basically everyone is familiar with this principle. It is often described as species being in constant fight with each other over a limited amount of resources. In this fight only those species that outperform others will have access to the limited amount of resources, whence will be able to pass on its reproductive code to next generations causing the selection.

We would like to generalize this principle to the setting of computations. This leads us to what we call the principle of Generalized Natural Selection: 

\begin{quote}
{\bf $\mathbf{GNS}$:} In nature, computational processes of high computational sophistication are more likely to maintain/abide than processes of lower computational sophistication provided that sufficiently many resources are around to sustain the processes.
\end{quote}

If one sustains the view that all natural processes can be viewed as computational ones, this generalization is readily made. For a computation, to be executed, it needs access to the three main resources  space, matter, and time. If now one computation outperforms the other, it will win the battle over access to the limited resources and abide. What does outperform mean in this context? 

Say we have two neighboring processes $\Pi_1$ and $\Pi_2$ that both need resources to be executed. Thus, $\Pi_1$ and $\Pi_2$ will interfere with each other. Stability of a process is thus certainly a requirement for survival. Moreover, if $\Pi_1$ can incorporate, or short-cut $\Pi_2$ it can actually use $\Pi_2$ for its survival. As an analogy we mention a monkey that can predict and thereby use the behavior of an ant by inserting a stick into an ant colony waiting for ants to climb on the stick so that the monkey can eat the ants by pulling the stick out again. 

A generalization of incorporating, or short-cutting is given by the notion of simulation that we have given above. Thus, if $\Pi_1$ can simulate $\Pi_2$, it is more likely to survive. In other words, processes that are of higher computational sophistication are likely to outperform and survive processes of lower computational sophistication. In particular, if the process $\Pi_1$ is universal, it can simulate any other process $\Pi_2$ and thus is likely to use or incorporate any such process $\Pi_2$.

Of course this is merely a heuristic argument or an analogy rather than a conclusive argument for the $\mathbf{GNS}$ principle. One can think of experimental evidence where universal automata in the spirit of the Game of Life are run next to and interacting with automata that generate regular or repetitive patterns to see if, indeed, the more complex automata are more stable than the repetitive ones. In setting up such experiments, much care needs to be taken to not run into hard philosophical problems of ontological nature like the question "what are the defining properties of a particular process". One can think of similar questions about a tree without leaves still being a tree etc. In particular, it seems more sensible to focus on some particular features, like for example entropy or other complexity measures. We will take up these considerations in more detail in Section \ref{section:TestingGNS}.

Of course, one cannot expect that experiments and circumstantial evidence can substitute or prove the principle.
A more detailed discussion of the principle can be found in \cite{JoostenGNS}

Just like the theory of the selfish gene 
(see \cite{SelfishGene}) shifted the scale on which natural selection was to be considered, now $\mathbf{GNS}$ is an even more drastic proposal and natural selection can be perceived to occur already on the lowest possible level: individual small-scale computational processes.

In \cite{JoostenGNS} it was noted that under some reasonable circumstances we may see $\mathbf{GNS}$ as a consequence of $\mathbf{PCE}$. 
However, $\mathbf{GNS}$ only talks about computational processes in nature and not in full generality about computational processes either artificial or natural as was the case in $\mathbf{PCE}$.  Thus we cannot expect that $\mathbf{CT}+  \mathbf{GNS}$ is actually equivalent to $\mathbf{PCE}$. However, if we restrict $\mathbf{PCE}$ to talk only about processes in nature, let us denote this by $\mathbf{PCE}'$, then we do argue that we can expect a correspondence. That is:
\[
{\mathbf{PCE'}} \ 
\approx
\ \mathbf{CT}\  + \ \mathbf{GNS}.
\] 
But ${\mathbf{PCE'}}$ tells us that almost all computational processes in nature are either simple or universal. If we have ${\mathbf{GNS}}$ we find that more sophisticated processes will outperform simpler ones and ${\mathbf{CT}}$ gives us an attainable maximum. Thus the combination of them would yield that in the limit all processes end up being complex. The question then arises, where do simple processes come from? (Normally, the question is where do complex processes come from, but in the formal setting of $\mathbf{CT}  +\mathbf{GNS}$ it is the simple processes that are in need of further explanation.)

Simple processes in nature often have various symmetries.  As we have argued above these symmetries are readily broken when a simple system interacts with a more complex one resulting in the simple system being absorbed in the more complex one. We see two main forces that favor simple systems. 

The first driving force is what we may call \textit{cooling down}. For example, temperature/energy going down, or material resources growing scarce. If these resources are not available, the complex computations cannot continue their course, breaking down and resulting in less complex systems.

A second driving force may be referred to as \textit{scaling} and invokes mechanisms like the Central Limit Theorem. The Central Limit Theorem is a phenomenon that creates symmetry by repeating a process with stochastic outcome a large number of times yielding the well-known Gaussian distribution. Thus the scale (number of repetitions) of the process determines the amount of symmetry that is built up by phenomena that invoke the Central Limit Theorem. 

In analogy, we can mention that whilst various universal processes that are executed at cell level, a tree by itself can hardly be called a universal computational process.

In the above, we have identified a driving force that creates complexity ($\mathbf{GNS}$) and two driving forces that creates simplicity: cooling down and scaling. In the light of these two opposite forces we can restate $\mathbf{PCE'}$ as saying that simplicity and universality are the two main attractors of these interacting forces. 

Note that we deliberately do not speak of an equivalence between $\mathbf{PCE'}$ and $\mathbf{CT}\  + \ \mathbf{GNS}$. Rather we speak of a correspondence. It is like when modeling the movement of a weight on a spring on earth. The main driving forces in this movement are gravitation and the tension of the spring. However, this does not fully determine a final equilibrium if we do not enter in more details taking into account friction and the like. It is in the same spirit that we should interpret the above mentioned correspondence.



\section{Complexity, Evolution and our Cognitive Toolkit}\label{section:CognitionAndComplexity}

Fodor has postulated a principle concerning our language. It says that (see \cite{FrideonsFodor}) the structure and vocabulary of our language is such that it is efficient in describing our world and dealing with the frame problem. The frame problem is an important problem in artificial intelligence which deals with the problem how to describe the world in an efficient way so that after a change in the state of affairs no entirely new description of the world is needed. See for example \cite{Shanahan1997}.

In particular, Fodor considers particles that can be either frigeons or nonfrigeons. A particle is a frigeon if Fodors refrigerator happens to stand open and otherwise it is a nonfrigeon. It is clear that we can perfectly well define such concepts and words. However, the mere availability of these concepts will not help us understand the world better. Nor are we likely to be able to act better in a competitive setting by having access to these concepts. And what is even worse, our description of the world becomes very cumbersome if we take these concepts into account. In particular of course at moments when Fodor's refrigerator is either opened or closed.

Based on this thought experiment Fodor posed the thesis that our language --an essential part of our cognitive toolkit-- has evolved in such a way to efficiently describe the world and the important changes occurring therein.

On a similar page, we would like to suggest that our cognitive toolkit has evolved over the course of time so that it best deals with the processes it needs to deal with. Now, by $\mathbf{PCE}$ these processes are either universal or very simple. Thus, it seems to make sense in terms of evolution to have a cognitive toolkit that is well-suited to deal with just two kinds of processes: the very simple ones and the universal ones.

Taking these considerations into account, it can well be conceved that there actually are computational processes out there that violate $\mathbf{PCE}$ but firstly, by $\mathbf{GNS}$ these processes will be very scarce and secondly, even if they are out there, our cognitive toolkit is just not well-equipped enough to deal with them.

Actually, throughout mathematics and mathematical logic there are various indications present that seem to substantiate the claim that indeed many of our most commonly used intellectual and cognitive tools within these fields, although rather sophisticated, all fall in one of few classes of operational strength. In this paper we have already seen that it is very hard to get sets that are not computationally universal. In \cite{JoostenGNS} we gave some more examples to this same phenomenon. 

In this setting we would also like to mention the program of reverse mathematics (see \cite{Simpson1997}). Reverse mathematics tries to gauge the logical strength of important mathematical theorems. One starts out with some weak base theory $T_0$. Next, one considers some important mathematical theorem $\tau$. These are typically mathematical theorems that are frequently used by the mathematical community. 

We mention here some examples of such theorems without further reference, context or proof. They just serve to give the flavor of the kind of theorems considered: 
\begin{itemize}

\item
Every countable commutative ring has a prime ideal; 

\item
A continuous real function on the closed unit interval is Riemann integrable;

\item
Uniqueness of algebraic closure (of a countable field);

\item
G\"odel's completeness theorem: a formula $\varphi$ in a countable language is provable from a set $\Gamma$  of assumptions in that same language, if and only if $\varphi$ is true in every model where all of $\Gamma$ is true.

\end{itemize}

As said, we do not want to go into the details of these theorems. They merely serve the purpose of illustrating what kind of theorems are considered and how wildly divers the scope of these different theorems are. The next step in the recursive mathematics project is to consider the system $T_0 + \tau$, that is, the base system together with one of those particular mathematical theorems. We call two such systems $T_0 + \tau$ and $T_0 + \tau'$ equivalent and write $T_0 + \tau \ \equiv \ T_0 + \tau'$, if they prove exactly the same set of theorems, that is,
\[
T_0 + \tau \vdash \varphi \ \ \Longleftrightarrow \ \ T_0 + \tau' \vdash \varphi
\]
for any formula $\varphi$. It turns out that almost all important mathematical theorems fall into one of five equivalent systems. That is to say, if you take six arbitrary important mathematical theorems $\{ \tau_i \mid i\in \{1,\ldots , 6\}\}$, almost surely you will have that $T_0 +\tau_i \ \equiv \  T_0 + \tau_j$ for some $i\neq j$. We think that this is an important indication of the fact that our intellectual/cognitive toolkit is designed in such a way as to efficiently/naturally recognize, and deal with a limited set of problems that are most useful to us in our daily life and fight for survival. In particular we mention that all the above mentioned examples of important mathematical theorems are equivalent over some natural base theory $T_0$.

\section{On testing the Generalized Natural Selection principle}\label{section:TestingGNS}

In this final section we shall address the question on how to test the principle of Generalized Natural Selection $\mathbf{GNS}$ as put forward in \cite{JoostenGNS} and discussed here in Section \ref{section:ComplexityAndEvolution}. As mentioned before, such tests can never substitute a full proof. Rather they can merely supply ``circumstantial evidence'' in favor of or against the principle. Let us start out by pointing out some subtleties underlying $\mathbf{GNS}$.

\subsection{General observations}\label{section:GeneralObservations} 

The principle $\mathbf{GNS}$ tells us that complex processes are more likely to be more successful than others and thus more likely to ``survive''. Let us recall the exact formulation of $\mathbf{GNS}$ and make some general observations that should always be taken into account when studying it.

\begin{quote}
{\bf $\mathbf{GNS}$:} In nature, computational processes of high computational sophistication are more likely to maintain/abide than processes of lower computational sophistication provided that sufficiently many resources are around to sustain the processes.
\end{quote}

\noindent
In this formulation we see the following difficulties naturally emerge.

\begin{enumerate}
\item
The first, most natural, and most fundamental question is ``what is determining the identity of a process''. For example, suppose some process $\Pi$ undergoes some minimal change, should we still call it the same process $\Pi$ after that minimal change? The same sort of question arises in all kinds of sciences: ``is a human being without limbs still a human being?'' or ``when is a particular cloud a Nimbo Cumulus?''. It turns out even to be difficult to classify life within very broad categories like ``animal'' versus ``plant" etc.

Naturally the question is related to deep philosophical questions relating, amongst others, to issues like fuzziness (see e.g., \cite{Zadeh1996}) in particular and the Sorites paradox more in general (\cite{Keefe2000}). The Sorites paradox deals with questions like ``how many trees should there be to form a forest, and what happens if I would cut one tree".

When trying to make quantified statements about $\mathbf{GNS}$ one should always first isolate a well defined entity that substitutes either ``process" or some essential property representing the process. In the paradigm of \emph{The selfish gene} this is easy but in the paradigm of $\mathbf{GNS}$ this is not.

\item
A second fundamental question is concerning what is meant by complexity and in particular how to measure it. As mentioned before, there are various essentially different definitions throughout the literature and we have put forward our own proposal here in Section \ref{section:ComplexityAndComputation}.

\item
We think that the two problems mentioned so far are the more serious ones. Minor but not less fundamental problems arise in also defining the other mentioned concepts. Thus, how should we specify probability when saying that one process is more likely to maintain/abide than another. What is exactly understood by interaction, etc.
\end{enumerate}

\subsection{Mathematical analysis}
In principle one could try to formulate $\mathbf{GNS}$ in a fully formalized setting and then try to \emph{prove} $\mathbf{GNS}$ as a theorem within that formal setting. In doing so all above mentioned points/problems should be taken into account. We shall shortly see how many choices such an analysis entails. That naturally raises the questions on how natural these choices are and in how much the final analysis says something about the physical reality at all.

For example, one could identify a process by a set, or better, by a Turing degree $X$. This would be a first choice. In a next choice one has to define some mathematical operation $\oplus$ between two degrees that models the notion of interaction between two processes aka degrees. Thus, the outcome of two processes $X$ and $Y$ that interacted would be denoted and computed by $X\oplus Y$.

Subsequently, we can answer the question whether $X \leq X\oplus Y$ and  $Y \leq X\oplus Y$. However, if we wish to say something on how likely it is that either $X \leq X\oplus Y$ or $Y \leq X\oplus Y$ we need to make yet more choices like introducing some probabilistic tools on the space of degrees between $\emptyset$ and $\emptyset'$.

\subsection{Testing in the Laboratory}
Instead of mathematical modeling, one can also try to isolate some real-world processes that can be considered naturally as computational ones and have them interact and run in a laboratory setting. Also here, all the difficulties as discussed in Section \ref{section:GeneralObservations} will manifest themselves in this setting. 

In particular we will have to decide on the identity determining aspects of the processes involved. Moreover we have to decide on the measure of complexity that is to be applied to these processes. Probably that is the harder and more arbitrary task in this setting. 

When working with living organisms some care has to be taken as to prevent that we are just testing the well-established principle of natural selection instead of $\mathbf{GNS}$.

\subsection{Computer simulations}
In this final section we shall discuss a possible approach to test $\mathbf{GNS}$ via computer simulations. Again we should settle upon choices for the modeling problems as posed in \ref{section:GeneralObservations}.

Just as with the mathematical modeling, we would like to stay as close as possible to the physical reality in our computer simulations. Due to the inherent parallel nature of physical reality (all goes on at the same time) and due to the locality of causality it seems a good idea to simulate parts of reality by cellular automata (CA). 

\begin{figure}
  \centering
  \includegraphics[width=10cm]{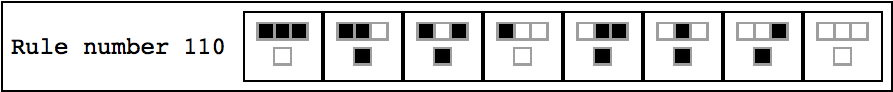}
  \caption{Definition of Rule 110}
  \label{Figure:rule110Definition}
\end{figure}

For the sake of a simple presentation let us briefly recall the definition of one of the simplest CAs: a one-dimensional CA with radius 1 and two symbols. We shall depict the two different symbols by black and white respectively. Our CA acts on a one dimensional tape of discrete cells that extend infinitely both to the left and to the right. In CAs, time evolution is modeled by discrete time steps. An initial condition is given by telling what cell is of what color. The color of each cell will evolve over time. Basically our CA is just a look-up table with a rule how to compute the color of a particular cell at a next time step depending on its current color and the color of its two direct neighbors. An example of such a CA is depicted in Figure \ref{Figure:rule110Definition}. 

The rule numbering is according to a numbering scheme as presented in \cite{NKS} but not really relevant for the current presentation. Thus, for example, if a cell was white at time $t$ and both its neighbors were black at time $t$, then at the next step $t+1$, the cell will turn black according to the defining look-up table of Rule 110.

As in \cite{NKS} we depict the consecutive tape configurations from top to down. Thus, for example, if we start out with just a single black cell on an otherwise white tape, Rule 110 will give us the famous evolution as depicted in Figure \ref{Figure:rule110Evolution} below. 

\begin{figure}
  \centering
  \includegraphics[width=8.45cm]{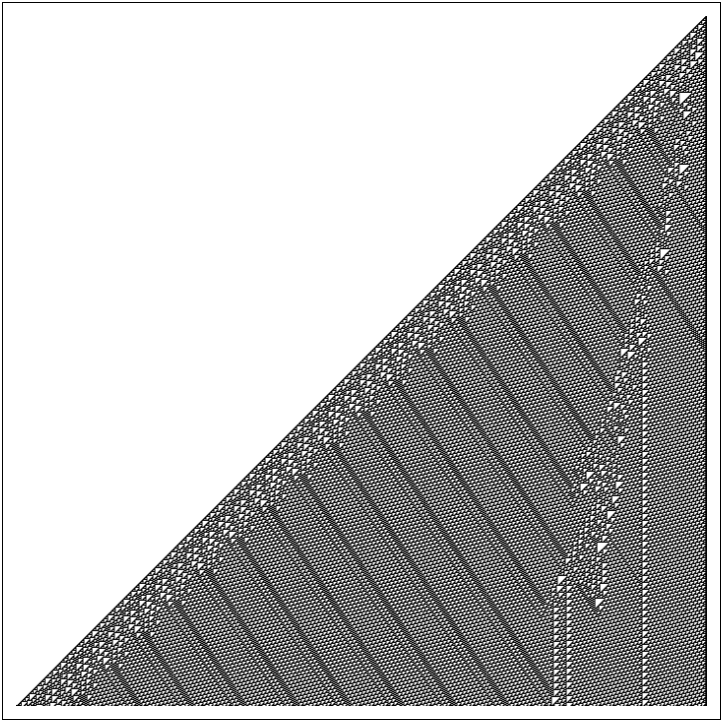}
  \caption{Evolution of Rule 110 starting with just one black cell and computed only for 700 steps. Figure generated with \emph{Mathematica}.}
  \label{Figure:rule110Evolution}
\end{figure}

It is evident from Figure \ref{Figure:rule110Evolution} that complex behavior can already occur in these simple automata. As a matter of fact, it is know that Rule 110 is universal in that it can --in some sense-- emulate any other computational process. It is easy to see how CA can be generalized to more symbols, more dimensions and larger radii (taking more neighbors into account).

To come back to our test of $\mathbf{GNS}$ and in the setting of CAs, how should we model processes? Should a process be modeled by a particular CA? And if so, how should interaction be modeled? We think it is more natural to model a process by an initial condition. Let us briefly explain why.

We have reasoned before that there is a strong analogy between physical reality and CAs. Each cell in a CA with its respective symbol can be seen as a particular property of physical reality at some particular locus or region. The interaction between these properties at these regions are governed by the same laws of nature everywhere throughout the universe. At least it is generally believed to be the case that the laws of physics are the same throughout the universe. 

One could not wish to adhere to this believe and keep the possibility open that somewhere far away in extreme circumstances --for example close to a black hole-- the laws of physics do change. However, it still seems reasonable to expect the laws of nature to be at least locally stable. And as we are interested on interacting processes it is mainly local interaction that we are interested in. Thus, if we have a CA simulating physical reality, it should be the same CA at every part of the tape. Moreover, if we wish to simulate reality in which universality clearly occurs, we better start out with a universal CA\footnote{This requirement can be relaxed as a universal CA can mimic any other CA too.}. Pushing the analogy further we are lead to accept that processes correspond to the configuration of our symbols evolving over time. 

It is in this setting that the question about the defining properties of a particular process becomes very hard. Thus, in simulations using CAs it seems more fruitful to focus on particular features of a process rather than to find a set of defining properties that sharply tells us what a particular process is. For the moment we shall not address the issue of limited resources. 

So now that we have identified a process with an initial condition the problem of finding a suitable definition of complexity becomes clearly defined. Remember that we propose to work first with one-dimensional CAs with just two symbols. Thus, a process is nothing but a string developing over time for which there are suitable and effective complexity measures defined (see \cite{DelahayeZenil2012} or \cite{Zenil2010}). By brute force simulations one can now try to quantify how likely it is that the more complex processes maintain/abide. 

Even if these simulations could provide circumstantial evidence in favor of $\mathbf{GNS}$, one still has to be very careful in how to interpret the repercussions of these simulations on the physical reality. However, positive outcomes of such simulations, experiments, or theorems will certainly help gain credibility of $\mathbf{GNS}$. Further credibility could be obtained by applications of $\mathbf{GNS}$ in related theoretical frameworks and only time will tell if these are to be found or not. 

\section*{Acknowledgments}
I would like to thank Barry Cooper and Hector Zenil for fruitful discussions on the subject.

\end{document}